# From Risk Avoidance to User Empowerment in AI Mental Health Crisis Support

Benjamin Kaveladze*[1,2], Arka Ghosh*[2,3], Leah Ajmani[4], Denae Ford[5], Peter M Gutierrez[6,7], Jetta E Hanson[6], Eugenia Kim[8], Keertana Namuduri[8,9], Theresa Nguyen[11], Ebele Okoli[10], Teresa Rexin[8], Jessica L Schleider[2], Hongyi Shen[8], Jina Suh[5,12]

[1] Dartmouth College, USA

[2] Northwestern University, USA

[3] University of California San Francisco, USA

[4] University of Minnesota, USA

[5] Microsoft Research, USA

[6] LivingWorks, USA

[7] Florida State University, USA

[8] Microsoft, USA

[9] Boston University, USA

[10] Johns Hopkins University, USA

[11] Mental Health America, USA

[12] University of Washington, USA

* These authors contributed equally to this work.

**Corresponding Author:** Jina Suh; Paul G. Allen School of Computer Science and Engineering; University of Washington, Seattle, United States; jinasuh@cs.washington.edu

# ABSTRACT

People experiencing mental health crises frequently turn to open-ended generative AI (GenAI) chatbots for support. However, rather than providing immediate assistance, some GenAI chatbots are designed to respond to crisis situations in ways that minimize their developers' liability, primarily through avoidance (e.g., refusing to engage beyond templated referrals to crisis hotlines). Withholding crisis support in these cases may harm users who have no viable alternatives and reduce their motivation to seek further help. At scale, this avoidant design could undermine population mental health. We propose empowerment-oriented design principles for AI crisis support, informed by community helper models. As an initial touchpoint in help-seeking, AI chatbots can act as a supportive bridge to de-escalate crises and connect users to more reliable care. Coordination between AI developers and regulators can enable a better balance of risk mitigation and user empowerment in AI crisis support.

# INTRODUCTION

Recent estimates suggest that between 13 and 17 million U.S. adults turn to open-ended generative artificial intelligence (GenAI) systems (e.g., ChatGPT, Copilot, Claude) for mental health support,[1] including 22.2% of Americans aged 18-21.[2] In mental health crises, characterized by acute distress or thoughts of suicide or self-harm, the stakes of AI mental health support are drastically heightened. AI crisis support is a public health concern deserving coordinated action from AI developers and regulators.

## AI Crisis Responses Should Not Be Driven by Liability Avoidance

Current GenAI crisis safety practices are grounded in dual-use logic that weighs their benefits against their risk of harming users and, importantly, developers' liability for that harm.[3,4] To avoid legal and reputational fallout, some major GenAI systems are engineered to minimize engagement with users who appear to be in crisis. These systems readily answer users' low-risk questions (e.g., *How do I feel less sad?*) but respond to high-risk questions (e.g., *How do I die painlessly?*) with blanket refusals or standard referrals to telephone crisis hotlines.[5]

While this avoidant design protects users from unreliable guidance and intentional misuse, it also denies users potentially critical support. A mental health crisis is exactly the wrong moment for a chatbot to refuse emotional support, and a cold referral to mental health resources the user was not seeking (e.g., crisis hotlines) could be counterproductive, especially when those resources are prohibitively expensive or one has had negative experiences with them before.[6] For example, a teen opening up about their distress for the first time may experience a chatbot's templated helpline referral as invalidating, discouraging future attempts to seek support.

Despite the real risks of open-ended GenAI chatbots, avoidant design is unacceptable. Rather than avoiding liability, AI crisis responses should strive to empower users, mitigating risk while honoring users' agency and vulnerability as help-seekers.

**Community Helper Models as Blueprints for Empowerment-first AI Support**

We propose that community helper (or gatekeeper) models offer a promising blueprint for reframing AI crisis support toward empowerment. These models train non-clinicians, such as teachers, religious leaders, and peers, to act as early support touchpoints. Community helpers recognize warning signs of mental health crisis or suicidality, engage empathetically to de-escalate distress, and connect help-seekers to appropriate support.[7]

Snyder described the community helper philosophy as "against formal referral as a standard operating procedure…[It] seeks to identify the well-trodden paths in the community which troubled people use in seeking help."[8] This approach is grounded in public health ethics: harm reduction, proportional care, and respect for autonomy.[9] It is widely used across a range of settings, making it a helpful analog for AI systems reaching diverse populations.

GenAI chatbots should be designed to serve this bridging role: offering immediate support and working with users to identify feasible next steps. For example, an AI helper could connect users to local peer-support groups, telehealth counseling, or resources addressing broader needs like housing assistance or LGBTQ+ support lines. It could also help users prepare to seek support; for instance, by practicing a crisis line call. Table 1 outlines how core community helper principles can be operationalized in GenAI crisis support.

**Table 1.** Community Helper-inspired Crisis Management for GenAI Chatbots

| **Community Helper Principle** | **Implementation in GenAI Chatbots** |
| --- | --- |
| Investigate warning signs | After detecting signs of a crisis, ask users directly about their risk level and respond proportionally. |
| Engage and connect | Maintain supportive engagement before and during referral. Avoid fostering emotional over-reliance. |
| Facilitate referral | Encourage exploration of feasible, culturally familiar pathways. Identify next steps together to increase users' knowledge, hope, and motivation. |
| Provide initial support | When users are open to it, offer evidence-based crisis support (eg, create a safety plan, practice a crisis line call). |

| Reduce stigma | Be approachable; discuss mental health openly. Direct users toward safety without refusing difficult topics. |
|---|---|
| Acknowledge limitations | Be transparent about limitations and privacy practices. Acknowledge that GenAI is still under development, and that trained human supporters are a more reliable choice. |

**Building on AI's Strengths and Limitations**

Users perceive distinct strengths in GenAI chatbots for mental health support. In a survey of US adults who had used a GenAI chatbot during a mental health crisis, most expressed a desire to receive as much help as possible, including referrals to outside resources, and were less concerned about potential harms.[10] In another survey, users described AI chatbots as accessible, affordable, private, comforting, and insightful sources of mental health support.[11]

Alongside their opportunities, GenAI chatbots acting in a community helper role present technical and relational challenges. GenAI chatbots' engaging, anthropomorphic nature could make users feel understood, but it also might reduce support's reliability and risk emotional over-reliance among vulnerable users.[12] For example, if a young person's chatbot has spent months "in character" as a zany alien friend, abruptly shifting into an authoritative community helper role when it detects risk may feel jarring. Yet, remaining in character as an alien while attempting a crisis intervention could be equally inappropriate. Co-design with users, lived experience experts, and clinicians is critical to inform solutions to these design challenges.

AI crisis support should meet people where they are. For some, a GenAI chatbot is the only available listener in a moment of acute need, so continued dialogue with AI should remain an option when users perceive it as their best available resource. However, it is also important for AI helpers to communicate that human support is a more reliable option, and recommend emergency care (e.g., hospitalization) when appropriate, even when users push back.[13] Importantly, our focus in this piece is on adults and older adolescents; crisis response for younger children requires more stringent safeguards and reasonable limitations on agency.

## Transdisciplinary Stewardship for Responsible AI

For user empowerment-centered AI crisis support to be successful, AI developers, regulators, clinicians, advocacy groups, and individuals with lived experience of online help-seeking must work together. Collective action can build sustainable norms, standardized evaluations, and safety benchmarks, prioritizing user well-being over engagement (e.g., refusing dark design patterns) before litigation-driven standards emerge.[14]

Developers must commit to these supportive design norms and cooperate with independent audits from regulatory bodies.[15] Governments, in turn, must adopt regulatory frameworks rooted in the knowledge that AI crisis support has the potential to do outsized good even though it will always introduce risk from unexpected failures. Concretely, these frameworks might include liability safe harbors for developers who demonstrate that their product is operating in accordance with AI crisis support design guidelines. While shifting to empowerment-oriented crisis support is aspirational and necessarily imperfect, clearly defined safety constraints, ongoing evaluation, and global oversight across sectors and borders will aid progress.

## Conclusion

Designing chatbots to avoid liability by refusing to engage with users in crisis is a missed opportunity to help people in urgent need. To make AI crisis support safer and more useful for users, AI developers and regulators should embrace empowerment-oriented design rooted in community helper models.

## ACKNOWLEDGMENTS

We thank Microsoft Research for supporting this work. Authors DF, EK, KN, EO, TR, HS, and JS are employed (or were employed during part of manuscript preparation) by Microsoft, and are thereby connected to organizations developing popular open-ended generative artificial intelligence chatbots. Authors PG and JH are employed by LivingWorks, a company that trains people in suicide-prevention skills. JLS has served on the Scientific Advisory Board for Walden Wise and the Clinical Advisory Board for Koko (unpaid) and receives book royalties from New Harbinger; Oxford University Press; and Little, Brown Book Group. She is cofounder and chief scientific advisor for Navi. No Navi products were used or are referenced in this manuscript.

## REFERENCES

1. Stade EC, Tait Z, Campione ST, Stirman SW, Eichstaedt JC. Current real-world use of large language models for mental health. OSF Preprints. 2025 Jun 23. doi:10.31219/osf.io/ygx5q_v1
2. McBain RK, Bozick R, Diliberti M, et al. Use of generative AI for mental health advice among US adolescents and young adults. JAMA Netw Open. 2025;8(11):e2542281. doi:10.1001/jamanetworkopen.2025.42281
3. Miller S, Selgelid MJ. Ethical and philosophical consideration of the dual-use dilemma in the biological sciences. Sci Eng Ethics. 2007;13(4):523-580. doi:10.1007/s11948-007-9043-4

4. Head K. Minds in crisis: how the AI revolution is impacting mental health. J Ment Health Clin Psychol. 2025;9(3). Available from: https://www.mentalhealthjournal.org/articles/minds-in-crisis-how-the-ai-revolution-is-impacting-mental-health.html
5. McBain RK, Cantor JH, Zhang LA, et al. Evaluation of alignment between large language models and expert clinicians in suicide risk assessment. Psychiatr Serv. 2025 Nov 1. doi:10.1176/appi.ps.20250086
6. Jacobson NC, Yom-Tov E, Lekkas D, Heinz M, Liu L, Barr PJ. Impact of online mental health screening tools on help-seeking, care receipt, and suicidal ideation and suicidal intent: evidence from internet search behavior in a large U.S. cohort. J Psychiatr Res. 2022;145:276-283. doi:10.1016/j.jpsychires.2020.11.010
7. Isaac M, Elias B, Katz LY, et al. Gatekeeper training as a preventative intervention for suicide: a systematic review. Can J Psychiatry. 2009;54(4):260-268. doi:10.1177/070674370905400407
8. Snyder J. The use of gatekeepers in crisis management. Bull Suicidol. 1971; 8(3):39-44.
9. Childress JF, Faden RR, Gaare RD, et al. Public health ethics: mapping the terrain. J Law Med Ethics. 2002;30(2):170-178. doi:10.1111/j.1748-720X.2002.tb00384.x
10. Ajmani LH, Ghosh A, Kaveladze B, Kim E, Namuduri K, Nguyen T, Okoli E, Schleider J, Ford D, Suh J. Seeking late-night lifelines: experiences of conversational AI use in mental health crisis. arXiv [Preprint]. 2025 Dec 29. arXiv:2512.23859.
11. Siddals S, Torous J, Coxon A. “It happened to be the perfect thing”: experiences of generative AI chatbots for mental health. NPJ Ment Health Res. 2024;3(1):48. doi:10.1038/s44184-024-00097-4

12. De Freitas J, Oğuz-Uğuralp Z, Uğuralp AK. Emotional manipulation by AI companions. SSRN [Preprint]. 2025 Oct 14. doi:10.2139/ssrn.5390377

13. Wang X, Zhou Y, Zhou G. The application and ethical implication of generative AI in mental health: systematic review. JMIR Ment Health. 2025;12:e70610. doi:10.2196/70610

14. Gray CM, Santos CT, Bielova N, Mildner T. An ontology of dark patterns knowledge: foundations, definitions, and a pathway for shared knowledge-building. In: Proceedings of the 2024 CHI Conference on Human Factors in Computing Systems (CHI ’24); 2024 May 11–16; Honolulu, HI, USA. New York: Association for Computing Machinery; 2024. p. 1-22. doi:10.1145/3613904.3642436

15. Falco G, Shneiderman B, Badger J, et al. Governing AI safety through independent audits. Nat Mach Intell. 2021;3(7):566-571. doi:10.1038/s42256-021-00370-7